\documentclass[conference]{IEEEtran}
\IEEEoverridecommandlockouts
\usepackage{cite}
\usepackage{amsmath,amssymb,amsfonts}
\usepackage{algorithmic}
\usepackage{graphicx}
\usepackage{textcomp}
\usepackage{xcolor}
\def\BibTeX{{\rm B\kern-.05em{\sc i\kern-.025em b}\kern-.08em
    T\kern-.1667em\lower.7ex\hbox{E}\kern-.125emX}}
\begin{document}

\title{Performance Reporting of Mathematical Library Installations with LAAB - An Overview
\thanks{This research was conducted using the resources of the High-Performance Computing Center North (HPC2N), Sweden. This work was supported by the eSSENCE Programme under the Swedish Government's Strategic Research Initiative, the J\"ulich Supercomputing Center at Forschungszentrum J\"ulich, Germany through the BMBF project 01-1H1-6013 AP6-NRW Anwenderunterst\"utzung SiVeGCS, and RWTH Aachen University, Germany through the DFG project IRTG-2379. }
% \thanks{Acknowledgements will be added here. Citations of the authors work related to this paper are marked ``[?]". }
}

\author{
    \IEEEauthorblockN{Aravind Sankaran}
    \IEEEauthorblockA{
        \textit{J\"ulich Supercomputing Center} \\
        \textit{Forschungszentrum J\"ulich, Germany} \\
        a.sankaran@fz-juelich.de
    }
    \and
    \IEEEauthorblockN{Paolo Bientinesi}
    \IEEEauthorblockA{
        \textit{Department of Computer Science} \\
        \textit{Ume{\aa} University, Sweden} \\
        pauldj@cs.umu.se
    }
}

% \author{
%     \IEEEauthorblockN{Author 1}
%     \IEEEauthorblockA{
%         \textit{Affiliation} \\
%         email
%     }
%     \and
%     \IEEEauthorblockN{Author 2}
%     \IEEEauthorblockA{
%         \textit{Affiliation} \\
%         email
%     }
% }

\maketitle

\begin{abstract}
%\p{to keep in mind: is "Reporting" the most suitable verb? To monitor, investigate, assess?}
% We present the Linear Algebra Aware Benchmarks (LAAB) framework, which facilitates performance reporting of mathematical-library installations on HPC systems. 
% We present the Linear Algebra Aware Benchmarks (LAAB) framework for systematically assessing and reporting the performance of mathematical library installations on HPC systems.
% Mathematical libraries provide interfaces for operations that form the computational building blocks of scientific applications. Reporting their performance is important not only for assessing the efficiency of scientific codes, but also for estimating compute-time requirements and preparing appropriate resource-allocation requests. 
% In this paper, we define four objectives for such performance reporting:
% 1) traceability, which associates each report with the exact mathematical library installation and execution configuration under evaluation; 2) compatibility, which enables the performance of operations provided by mathematical libraries to be related to the performance of higher-level scientific applications that use them; 3) reliability, which supports the interpretation of reports in the presence of measurement variability; and 4) accessibility, which ensures that reports, benchmark definitions, and relevant metadata are available for inspection and reproduction. We then present the design of LAAB and show how it addresses the challenges for each objective.

% 150 word concise abstract required for submission:
We present the Linear Algebra Aware Benchmarks (LAAB) framework for systematically assessing and reporting the performance of mathematical library installations on HPC systems. Mathematical libraries provide interfaces for operations that form the computational building blocks of scientific applications. Reporting their performance is important for assessing application efficiency, estimating compute-time requirements, and preparing resource-allocation requests. In this paper, we define four objectives for performance reporting: 1) traceability, linking each report to the exact library installation and execution settings; 2) compatibility, relating library-operation performance to higher-level scientific applications that use them; 3) reliability, supporting interpretation in the presence of measurement variability; and 4) accessibility, ensuring that reports, benchmark definitions, and relevant metadata are available for inspection and reproduction. We then present the design of LAAB and show how it addresses the challenges associated with these objectives.

%About the Author:
% Aravind Sankaran is a Research Fellow at the Jülich Supercomputing Centre (JSC), Forschungszentrum Jülich, Germany, working in the area of high-performance computing- application support and optimization. His research focuses on performance benchmarking and analysis of mathematical libraries and scientific applications on HPC systems. He develops methods and tools for reproducible benchmarking, performance reporting, and the interpretation of performance variability across software and hardware environments. He holds an M.Sc. in Simulation Sciences from RWTH Aachen University and is pursuing doctoral research in HPC application performance reporting and analysis.

\end{abstract}

\begin{IEEEkeywords}
Mathematical software benchmarks, Performance regression, Reproducible benchmarks.
\end{IEEEkeywords}

\section{Introduction}

A wide range of mathematical libraries, such as those based on Basic Linear Algebra Subprograms (BLAS) and Linear Algebra Package (LAPACK), exist to simplify scientific software development and to enhance both human productivity and computer efficiency~\cite{lawson_basic_1979}~\cite{angerson_lapack_1990}. 
These libraries provide essential building blocks for developing scientific software applications. Some examples of mathematical operations provided by these libraries include general matrix–matrix multiplication, LU, QR, and Cholesky factorisations, the solution of linear systems, and the computation of eigenvalues and eigenvectors. Because these operations are frequently used in scientific applications, developed and deployed on High-Performance Computing (HPC) systems, HPC sites typically provide installations of the corresponding libraries configured for the target hardware, eliminating the need for users to install and tune the libraries themselves.

Although mathematical libraries reduce the burden of implementing and optimising numerical algorithms from scratch, and substantial effort is devoted to maintaining and providing these libraries on HPC systems~\cite{geimer_modern_2014, gamblin_spack_2015, droge_eessi_2023}, the performance of the operations they implement is not generally reported. The reporting of performance offers several benefits. First, it enables HPC site personnel to identify performance regressions following software or system updates and to ensure that users continue to benefit from a well-optimised scientific computing environment. Second, it helps users select and use external libraries effectively, estimate the compute time and resource requirements of large-scale simulations, and prepare appropriate allocation requests. Third, it supports the technical evaluation of compute-time applications and assists evaluators in making grant and resource-allocation decisions.  To serve these objectives, we present the Linear Algebra Aware Benchmarks (LAAB) framework, which facilitates the performance reporting of mathematical libraries on HPC systems.

The remainder of this paper is organised as follows. Sec.~2 provides definitions. The objectives are listed in Sec.~3. Challenges and related works are discussed in Sec.4, and the design of LAAB is presented in Sec.5. Sec.~6 demonstrates an application of the framework through an experiment, and Sec.~7 concludes the paper and outlines future work.

\section{Definitions}

We present definitions and terminology used in this work.

\textit{\textbf{Mathematical library}}. We consider a \textit{mathematical library} as a piece of software that provides interfaces for one or more \textit{mathematical operations} which involve computations with vectors, matrices, and/or tensors. 
% \p{sounds like you want to restrict the focus to "linear algebra libraries"} \as{Not really. I don't think high-level libraries like PyTorch can be called as linear algebra libraries. But they involve computations with vecs, matrices, tensors..} \p{fine, but that's the message you send with the clause "which involve computations with vectors, matrices, and/or tensors". I wonder if/why that is needed. I guess it fits the storyline of your dissertation, but it's not at all clear the role that plays in this article. } \as{may be you are right.. but I also wonder if we should have two different definitions for the dissertation and the paper.. }

\textit{\textbf{Step}}. We define \textit{step} as a unit of computation performed using a given mathematical library. A step may consist of a single mathematical operation or a sequence of such operations. It may also include I/O and data movement, such as fetching input batches from storage, transferring data between memory spaces, or exchanging data between processes using MPI.

Here are some examples of a step. If the report targets OpenBLAS, then a step may correspond to individual BLAS or LAPACK routines, such as \texttt{dgemm} for double-precision matrix--matrix multiplication or \texttt{dpotrf} for double-precision Cholesky factorization. For higher-level software, a step may be domain-specific. For example, in Quantum ESPRESSO~\cite{giannozzi_quantum_2009}, a popular library for electronic-structure calculations, a step may correspond to a single iteration within a density functional theory workflow, such as a single k-point computation involving distributed complex matrix operations and the solution of an eigenvalue problem. In PyTorch~\cite{paszke_pytorch_2019}, a step may correspond to a single forward and backward pass of a transformer model including the I/O time required to fetch a batch of training data.

\textit{\textbf{Benchmark (of a step)}}. We define a \textit{benchmark} as the measure of the \textit{execution time} of a given step invoked through a mathematical library installation and executed on a specified number of processing elements (NPEs), such as threads, processes, cores, accelerators or nodes.

\textit{\textbf{Performance metric}}. We define \textit{performance metric} as a function that maps execution time from one or more repeated executions of a benchmark to a quantity that estimates a specific aspect of effectiveness of a step.

For example, if the number of floating-point operations (FLOPs) required to complete a step is known or can be estimated, then it is possible to report the rate at which FLOPs are executed (FLOPs/s). This performance metric estimates the computational effectiveness of the step. Similarly, the ratio of the execution time using one processing element to that using multiple processing elements is called the speedup. Speedup indicates how effectively a program uses additional resources relative to its own single-processing-element execution.

\textit{\textbf{Performance profile}}. A collection of performance metrics from one or more benchmarks associated with a common step is a \textit{performance profile}.

For example, consider a set of benchmarks in which a given step is executed using different numbers of processing elements. Note that scalability does not, in isolation, indicate absolute performance. Therefore, in addition to scalability, the achieved floating-point rate in FLOP/s is estimated for each benchmark. The collection of these metrics constitutes a performance profile.
As another example, consider benchmarks in which the same step is executed using different versions of a mathematical library. In this case, the performance profile may include a metric that assigns a rank to each library version.

%In this case, the performance profile may include a metric that computes a partial ranking~\cite{sankaran_ranking_2025} of the library versions.

A performance profile exists as a structured data, such as a Python object or a file containing the computed metrics.

\textit{\textbf{Performance report}}. 
A \textit{performance report} presents information from one or more performance profiles in a human-comprehensible format that enables performance assessment of the step under evaluation.
To prepare a performance report, implementation of visual elements, large language models for retrieval and presentation of relevant information from the profiles, and a user interface may be required. 

% \as{about your comment: is reporting the most suitable verb? The reason is I'm defining "reporting" here. I think reporting is more than just monitoring.. When you monitor, you collect information.. it does not capture the action of communicating the results..} \p{Agreed. Vice versa, I would not want readers to only interpret "reporting"  as communicating, missing the data collection. However, if it has to be a single word and the choice is between "reporting" and "monitoring", then I'd certainly take "reporting".}

\section{Objectives}

To support systematic performance reporting of mathematical library installations on HPC systems, we pursue the following four objectives:

\begin{enumerate}
    \item \textbf{\textit{Traceability}}. Multiple mathematical libraries may be available on an HPC system, often in several versions and build configurations optimised for different target hardware.  To enable meaningful performance assessment of a step implemented using a math library, each performance report should therefore be traceable to the exact library installation, including its version, build configuration, and target architecture. If required, the runtime configuration, including process-pinning settings, should also be recorded.

    \item \textbf{\textit{Compatibility}}. 
    Domain-specific software packages, such as PyTorch, Quantum ESPRESSO, and GROMACS, are typically built on top of core mathematical libraries, such as those providing BLAS and LAPACK implementations.
    To assess whether higher-level libraries make optimal use of the underlying core libraries, it should be possible to evaluate the mapping of high-level mathematical operations to the linear-algebra operations from the core libraries.

    \item \textbf{\textit{Reliability}}. 
    Run-to-run variability in performance measurements should be reported transparently, and performance assessment methods should be as robust as possible to such variations, thereby reducing the risk of drawing misleading conclusions.

    \item \textbf{\textit{Accessibility}}. 
    Performance reports, benchmark definitions, and relevant metadata should be available for inspection, comparison, and reproduction. Users should also be able to retrieve and rerun benchmarks with problem sizes and execution settings representative of their production workloads.
    
\end{enumerate}

\section{Challenges and Related Work}

We now discuss the challenges associated with achieving each objective and review the related work.

\textbf{\textit{Traceability}}. A typical HPC site consists of systems with multiple hardware architectures, such as different CPU generations and GPU accelerators. Each hardware target is often equipped with one or more toolchains, each typically consisting of a particular version of a base compiler (e.g., GCC, Intel oneAPI,  NVHPC), a threading library or programming model (e.g., OpenMP, POSIX threads, Intel TBB~\cite{dagum_openmp_1998, voss_pro_2019}), and an MPI implementation (e.g., Open MPI, Intel MPI, ParaStationMPI~\cite{gabriel_open_2004, suarez_modular_2022}).
Each math library is built on top of one or more of these toolchains, and this results in multiple software packages optimised for different targets with different toolchains while providing the same BLAS and LAPACK functionality, which are the foundations for both dense and sparse computations.

% Some examples of math libraries for CPU targets 
% \p{remove "targets"}
% include Intel MKL for Intel CPUs, NVPL for NVIDIA Grace CPUs, and AOCL for AMD CPUs. 
% \p{this sentence can be simplified: "libraries for CPUs includes Intel's MKL, NVIDIA's ... "  or "CPU math libraries include MKL (Intel), NVPL (NVIDIA), ..."}
% In this landscape, OpenBLAS and BLIS provide portable open-source implementations that can be built and tuned for new CPU architectures~\cite{goto_anatomy_2008, van_zee_blis_2015}. For GPU-accelerated workloads, corresponding libraries include cuBLAS for NVIDIA GPUs and rocBLAS for AMD GPUs. 
% \p{same comment: "include NVIDIA's cuBLAS and ..." or "include "cuBLAS (NVIDIA) and ..."}
% The required software also varies with the execution model. The distributed-memory workloads often require different libraries, such as ScaLAPACK for distributed linear algebra on CPUs and cuBLASMp for distributed GPU-accelerated BLAS operations~\cite{blackford_scalapack_1997}. SLATE~\cite{gates_slate_2019} targets distributed systems containing both CPUs and GPU.
% Beyond BLAS and LAPACK, specialised mathematical libraries support other operation classes, including FFTW~\cite{frigo_design_2005} for FFTs and PETSc~\cite{usdoe_office_of_science_sc_advanced_scientific_computing_research_ascr_petsctao_2025} for sparse linear algebra solvers. Corresponding GPU-oriented libraries include cuFFT, cuSOLVER  and their respective vendor-specific and distributed memory counterparts.

Some examples of libraries for CPU include Intel's MKL, NVIDIA's NVPL, and AMD's AOCL. In this landscape, OpenBLAS and BLIS provide portable open-source implementations that can be built and tuned for new CPU architectures~\cite{goto_anatomy_2008, van_zee_blis_2015}. Libraries for GPU-accelerated workloads include NVIDIA's cuBLAS and AMD's rocBLAS. The required software also varies with the execution model. The distributed-memory workloads often require different libraries, such as ScaLAPACK for distributed linear algebra on CPUs and NVIDIA's cuBLASMp for distributed GPU-accelerated BLAS operations~\cite{blackford_scalapack_1997}. SLATE~\cite{gates_slate_2019} targets distributed systems containing both CPUs and GPU.
Beyond BLAS and LAPACK, specialised math libraries support other operation classes, including FFTW~\cite{frigo_design_2005} for FFTs and PETSc~\cite{usdoe_office_of_science_sc_advanced_scientific_computing_research_ascr_petsctao_2025} for sparse linear algebra solvers. Corresponding GPU-oriented libraries include cuFFT, cuSOLVER  and their respective vendor-specific and distributed memory counterparts.

Apart from the maintenance of this large set of mathematical libraries, a further source of complexity is the large build configuration space associated with each library. That is, the same package may be built with different compiler flags, optimisation levels, instruction-set targets, and threading libraries. Exhaustively testing all possible combinations for all software is time consuming, which raises the question of which builds should be included in routine performance reporting.

In practice, many HPC centres use package-management frameworks such as EasyBuild~\cite{geimer_modern_2014}, Spack~\cite{gamblin_spack_2015}, or EESSI~\cite{droge_eessi_2023}, which encode tested installation procedures in build recipes that can be reused across similar systems.
However, having access to the deployed build recipes addresses only part of the problem.  The package management frameworks do not generally provide a structured mechanism for performance reporting of software installations. In this work, we use the available build recipes at a site as a practical basis for selecting representative software builds, and focus on performance reporting of those installations.

\textbf{\textit{Compatibility}}.
In recent years, user code bases have increasingly relied on high-level programming environments and software frameworks that are built on top of the core mathematical libraries discussed earlier, and it has been found that, in general, even when the core libraries achieve near-optimal performance on the target hardware, this does not necessarily imply that the users of higher-level libraries would fully exploit the capabilities of that the core libraries could have provided~\cite{psarras_linear_2022, sankaran_benchmarking_2022}. One reason is that high-level software packages may not expose interfaces that could take as input the structural information required to select specialised mathematical kernels; for example,
% \p{the example here seems to be a bit misplaced. You're going into details, when you said you would "discuss the related work and the challenges"} 
BLAS provides GEMM for general matrix–matrix multiplication of the form $AB$ and SYRK for operations of the form $AA^{T}$, where the result is symmetric. SYRK requires approximately half as many floating-point operations as GEMM and can, in principle, achieve up to twice the performance. However, high-level frameworks such as TensorFlow and PyTorch do not expose a dedicated SYRK interface, causing such operations to be expressed through general matrix multiplication and preventing this optimisation from being exploited~\cite{sankaran_benchmarking_2022}.
% \as{without that example, the reader would not be able to appreciate this related work statement, right?}. \p{see my comment in Sec.III} 
The same limitation applies to matrix multiplications involving triangular operands and to expressions that could be rearranged using the distributive law to reduce the FLOP count.

This problem of mapping a program written in a high-level language to an optimised sequence of core-library operations was investigated in~\cite{psarras_linear_2022} and referred to as the Linear Algebra Mapping Problem (LAMP). In this work, we separate performance reporting for core mathematical libraries from that for high-level, domain-specific libraries; for the latter, we aim to assess and report how effectively they solve LAMP.

%In this work, we separate the performance reporting of core mathematical libraries from that of high-level, domain-specific libraries and when assessing the performance of high-level libraries, we aim to report their capability to solve LAMP effectively.

% In this work, we separate the performance reporting of core mathematical libraries from that of high-level, domain-specific libraries.

% We address this compatibility issue by separating the performance reporting of core mathematical libraries from that of high-level, domain-specific libraries.
% Examples of the latter include benchmarks designed for PyTorch workloads, density functional theory workflows implemented in Quantum ESPRESSO, and similar application-level workloads. When the operations performed by high-level software can be traced to the corresponding operations in core libraries, it becomes possible to assess whether the underlying mathematical libraries are being used efficiently and with appropriate configuration settings.

\begin{figure*}[htbp]
    \centering
    \includegraphics[width=0.75\linewidth]{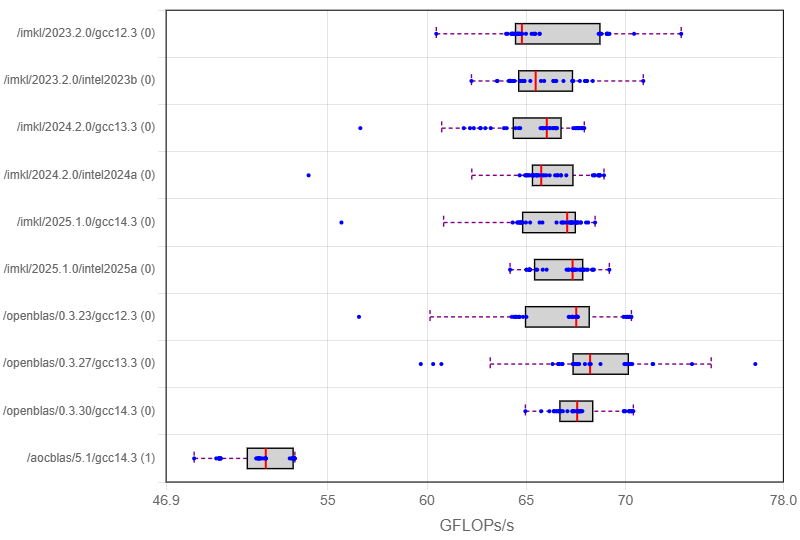}
    \caption{Single-core performance distributions of the BLAS \texttt{dgemm} kernel for 3000×3000 square matrices on an Intel Skylake CPU of the Kebnekaise
    HPC system. The box plots compare different versions and compiler-based installations of Intel MKL, OpenBLAS, and AOCL BLAS; the red line denotes the median performance in GFLOP/s. The Intel MKL and OpenBLAS installations exhibit overlapping distributions, suggesting comparable performance, whereas the AOCL installation shows a significantly different distribution, indicating noticeably worse performance.
    }
    \label{fig:gemm_bp}
\end{figure*}

\textbf{\textit{Reliability}}.
Performance measurements on HPC systems are often subject to run-to-run variations caused by factors such as shared network contention, memory hierarchy effects, cache state, system noise, and interference from other jobs~\cite{hunold_reproducible_2016, hoefler_scientific_2015, nikitenko_influence_2021, chen_statistical_2015, iakymchuk_improving_2011}.
In some cases, aggregate summaries, such as the mean, median or minimum across repeated runs, may be sufficient, for example when execution time is used only to verify whether specific algorithmic optimisations are enabled~\cite{psarras_linear_2022}. However, when comparing the same mathematical operation across different libraries, builds, or software versions, performance is expected to remain comparable unless a meaningful regression or improvement has occurred.
In such cases, execution-time-based analysis should account for measurement variability and identify ties instead of imposing an artificial strict ordering. However, deriving rankings with ties from noisy measurements is not straightforward, as multiple plausible rankings may exist, and the literature provides limited guidance on what constitutes a reasonable set of such rankings~\cite{sankaran_ranking_2025}.

% \p{same comment, again. It feels like this should be presented as motivation (in the intro), or moved somewhere else, or removed altogether.} \as{I had this before in the intro, but then intro will be big again if I put all the examples there.. I have revised this whole point a bit. Let me know if this example fits now} \p{see my comment in Sec.III}

For example,  consider the performance of double-precision matrix–matrix multiplication, commonly exposed through the \texttt{dgemm} kernel in BLAS libraries. We benchmarked this kernel using square matrices of size 3000$\times$3000 on a single core of an Intel Skylake CPU, comparing the implementation from multiple versions of OpenBLAS, Intel MKL, and AOCL BLAS built with different compiler versions and provided over the past two years on the Kebnekaise
HPC system at HPC2N, Sweden. The measured execution times are shown as box plots in Fig.~\ref{fig:gemm_bp}. The vertical axis identifies each library installation by library name, version, and toolchain. 
For example, \texttt{/imkl/2025.1.0/gcc14.3} denotes Intel MKL version 2025.1.0 math library used with GCC 14.3 compiler toolchain, whereas \texttt{/imkl/2025.1.0/intel2025a} denotes the same MKL version used with the Intel compiler toolchain. The horizontal axis reports the achieved performance in GFLOP/s. In each box plot, the box represents the interquartile interval, i.e., the range between the 25th and 75th percentile values, and the red line indicates the median.

We observe that AOCL BLAS 5.1 built with GCC 14.3, which was introduced only recently and was unavailable in earlier software stacks, exhibits a significant performance lag compared with the other installations. Identifying such regressions can trigger internal follow-up workflows in which the software team re-examines the build configuration, compiler flags, dependencies, and optimisation options and, where appropriate, reports the issue to the vendor to support improvements in future releases. 

However, ranking the installations solely by their median performance would assign a distinct rank to each version, disregarding the substantial overlap among the performance distributions of the remaining versions and thereby failing to clearly single out the under-performing installation. 
Allowing ties does not necessarily resolve this issue, as the resulting ranking may still be ambiguous. To account for this ambiguity and obtain a reasonable ranking with ties, we apply the partial-ranking methodology introduced in~\cite{sankaran_ranking_2025}.

% Allowing ties does not necessarily resolve this issue, as the resulting ranking may still be ambiguous~\cite{sankaran_ranking_2025}.

% A ranking that permits ties is not necessarily unique, and results in ambiguity~\cite{sankaran_ranking_2025}. To address this issue, we apply the partial-ranking methodology~\cite{sankaran_ranking_2025} which can address this issue. 

% to identify a set of reasonable rankings that allow for ties.

\textbf{\textit{Accessibility}}.
% The synthesis of a performance report for a given step from a set of benchmarks involves the following workflow.\p{"following workflow", but what follows does not read as a workflow. Maybe with some numbering? (i), (ii), ... or a) ..., b) ...}
% Each benchmark produces measurements including execution time of the selected kernel or high-level function call over repeated runs. Measurements resulting from multiple benchmarks are used to compute metrics and create performance profiles.
% In some cases, profiles generated on one environment must also be combined with profiles from benchmarks executed on other environments, resulting in a final profile that aggregates data across multiple set-ups to enable comparisons. These results must be updated continuously and automatically as additional benchmark runs become available for new versions of the libraries or toolchains.
% For reporting, the generated profiles should be translated into performance reports and made available in a user friendly format.
% To support both repeatability and reproducibility, users should be able to retrieve the corresponding benchmark definitions on the respective HPC systems, rerun them under the original conditions, or modify them as needed, with minimal additional dependencies apart from what is required to run the code.
The performance reporting for a given step from a set of benchmarks comprises the following workflow: (i) each benchmark is executed repeatedly to obtain measurements, including the execution time of the selected kernel or high-level function call; (ii) 
measurements from multiple benchmark runs are combined to compute metrics, construct comparative performance profiles, and update them automatically as new benchmark results become available, for example, for new library or toolchain versions; (iii) the resulting profiles are translated into performance reports and presented in a user-friendly format; and (iv) the corresponding benchmark definitions are made retrievable on the respective HPC systems, allowing users to rerun the benchmarks under the original conditions or modify them as required, with minimal dependencies beyond those needed to execute the benchmarked code.

Several existing tools contribute to this workflow, but each covers only a subset of the required capabilities. Frameworks such as JUBE~\cite{luhres_sebastian_flexible_2016, breuer_jube_2022} and ReFrame~\cite{karakasis_enabling_2020} primarily support step~(i) by providing environments for defining, executing, and managing the data generated by repeated benchmark runs. They also support parts of step~(iv) by retaining the information required to rerun a benchmark. However, they provide limited support for step~(ii), which requires measurements from multiple benchmark runs to be combined into performance metrics, and incorporated into performance profiles, and for step~(iii), which requires these profiles to be presented as user-oriented reports. The LLview~\cite{guimaraes_supporting_2026} framework can complement JUBE by supporting the reporting and visualisation of benchmark results, thereby addressing parts of step~(iii). However, it is not straightforward to re-use their user interface components to prepare performance reports that compare multiple profiles. Performance-analysis tools such as Vampir~\cite{resch_vampir_2008}, Scalasca~\cite{geimer_scalasca_2010} and Tau~\cite{shende_tau_2006} can collect, compute and visualise parts of the profiling information required in step~(ii) and (iii). They can perform cross-benchmark analysis, but do not provide the benchmark-management capabilities comparable to JUBE or ReFrame.
% but do not provide ranking methods for identifying performance regressions when performance distributions overlap.
% They also do not provide the benchmark-management capabilities comparable to JUBE or ReFrame.
%including the management of repeated benchmark runs and the retrieval of benchmark definitions for reproduction or modified reruns.
% They also do not manage benchmark execution or support the retrieval and rerunning of benchmark definitions required in steps~(i) and~(iv). 
% For this, they have to be complemented with tools such as JUBE or ReFrame. 
Consequently, to the best of our knowledge, no single existing tool encompasses the proposed workflow in its entirety.

\section{The Design of LAAB}
\label{sec:design}

\begin{figure*}[!p]
    \centering
    \includegraphics[
        width=\textwidth,
        height=0.9\textheight,
        keepaspectratio
    ]{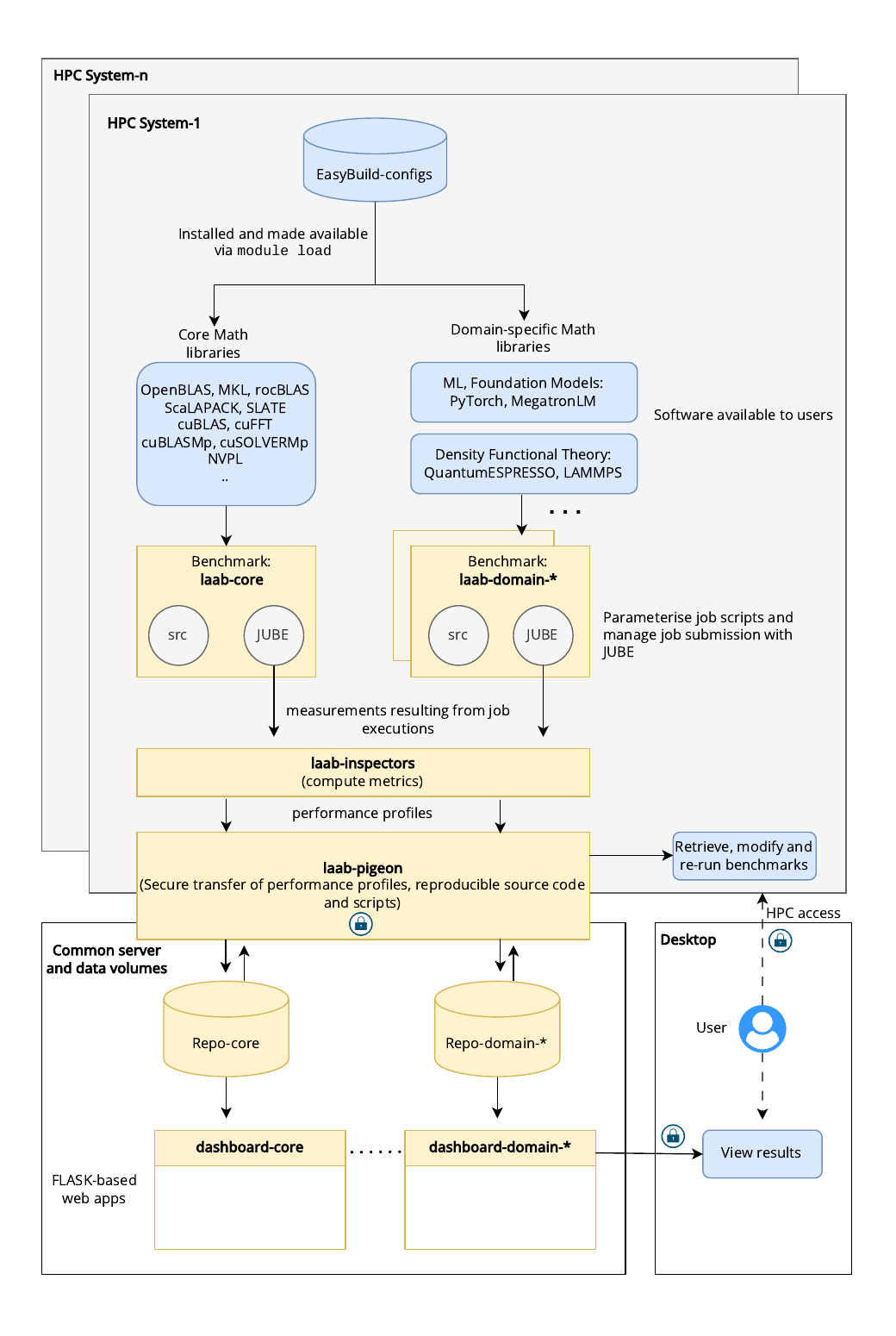}
    \caption{ Visual overview of the relations between the LAAB components described in Sec.~\ref{sec:design}. Yellow boxes denote components developed within LAAB, while blue boxes represent external software. The stacked HPC-system boxes indicate that benchmarks may be deployed on multiple systems. EasyBuild recipes identify the installations under test and JUBE parameterises and executes the jobs. Pigeon transfers performance profiles, source code, and scripts to common repositories used by the dashboards. The multiple dashboards illustrate the modular design of the user interfaces, which can be tailored to the varying reporting needs of the different libraries. The security symbols indicate site-provided security mechanisms through which LAAB benchmarks and performance profiles can be accessed.
    }
    \label{fig:laab}
\end{figure*}

The LAAB framework is expected to benefit three groups of beneficiaries: 1) HPC centre personnel responsible for providing, maintaining and monitoring the performance of mathematical libraries at the site, 2) HPC users who run scientific codes and aim to use the installed libraries efficiently, and 3) reviewers who evaluate scientific codes based on attainable performance, make allocation decisions and foresee user-support or research requirements. We envision that HPC centre personnel develop and maintain LAAB, prepare and execute benchmarks, and publish the reports together with the benchmark source code. Users and reviewers only inspect the reports and may retrieve and re-run relevant benchmarks with modified settings. 

In this section, we first describe the design of the LAAB performance-reporting workflow for developers implementing the framework, and then present a preview of the functionality available to users and reviewers.
A schematic overview is shown in Fig.~\ref{fig:laab}.
The components highlighted in yellow are part of the LAAB framework, while the remaining components interact with LAAB.

1) \textit{\textbf{Identification of software installation recipes (EasyBuild)}}. 
%Software installations available on the HPC system are generally made accessible to users through an environment module system, such as Lmod, and can be loaded using commands such as \texttt{module load}.
A requirement for LAAB is that the build recipe of the math library under evaluation is accessible, so that the build procedure, including configuration options, compiler flags, dependencies, and optimisation settings, can be identified.  Currently, we use EasyBuild recipes for this purpose. 
%However, the design is kept generic so that other software management frameworks, such as Spack, can also be supported.

In an EasyBuild-managed software environment, library installations are made accessible to users through an environment module system, such as Lmod, and can be loaded using the command \texttt{module load}.
As an example, consider loading the ScaLAPACK library on the Kebnekaise
HPC system, where software installations are managed using EasyBuild. ScaLAPACK is loaded with \texttt{module load GCC OpenMPI ScaLAPACK}. The modules \texttt{GCC} and \texttt{OpenMPI} constitute the toolchain dependency. For each loaded library, EasyBuild defines an environment variable of the form \texttt{\$EBROOT<NAME>}, which can be used to locate the corresponding EasyBuild recipe used for the installation. For ScaLAPACK, the recipe is found under \texttt{\$EBROOTSCALAPACK/easybuild/} and has the \texttt{.eb} extension. For this installation, the recipe file is \texttt{ScaLAPACK-2.2.2-gompi-2025b-fb.eb}. The file name encodes the main build information: \texttt{2.2.2} is the ScaLAPACK version, \texttt{gompi} indicates the GCC and OpenMPI toolchain dependency, and \texttt{2025b} identifies the versions of GCC and OpenMPI definied by the site for a specific year or stage. This file contains the configuration options, compiler flags, etc., used for the build.

2) \textit{\textbf{Benchmark creation (src).}} 
For each mathematical library, a set of representative steps and problem sizes is identified. Instrumentation is added around each step to measure its execution time; the measured values, together with a timestamp, are written to a log file, with one file generated per node irrespective of the number of processes.
% Each entry in this log file is referred to as an \textit{event}. 
Wherever available, an estimate of the FLOP count, the sizes of the input and output operands, and other information that can be derived from the function arguments are included in the log to support the estimation of performance metrics.

Benchmark source codes for core mathematical libraries and those of high-level software are maintained separately, allowing them to evolve independently and providing an extensible ecosystem for integrating custom benchmark code. In high-level software, language- or framework-level hooks in the high-level software can also be used to identify the underlying mathematical operations performed by core libraries.
%Benchmark source codes for core mathematical libraries and those of high-level software are maintained separately, allowing them to evolve independently and providing an extensible ecosystem for integrating custom benchmark code. Still, compatibility between the two categories must be ensured.
In~\cite{sankaran_benchmarking_2022},
we show an example of relating linear-algebra awareness of installations of popular machine-learning frameworks---TensorFlow and PyTorch---by comparing their performance with that of optimised implementations based on core mathematical libraries. 
%Alternatively, wherever possible, language- or framework-level hooks in the high-level software can also be used to identify the underlying mathematical operations performed by core libraries.

\begin{figure*}[!p]
    \centering
    \includegraphics[width=0.8\linewidth]{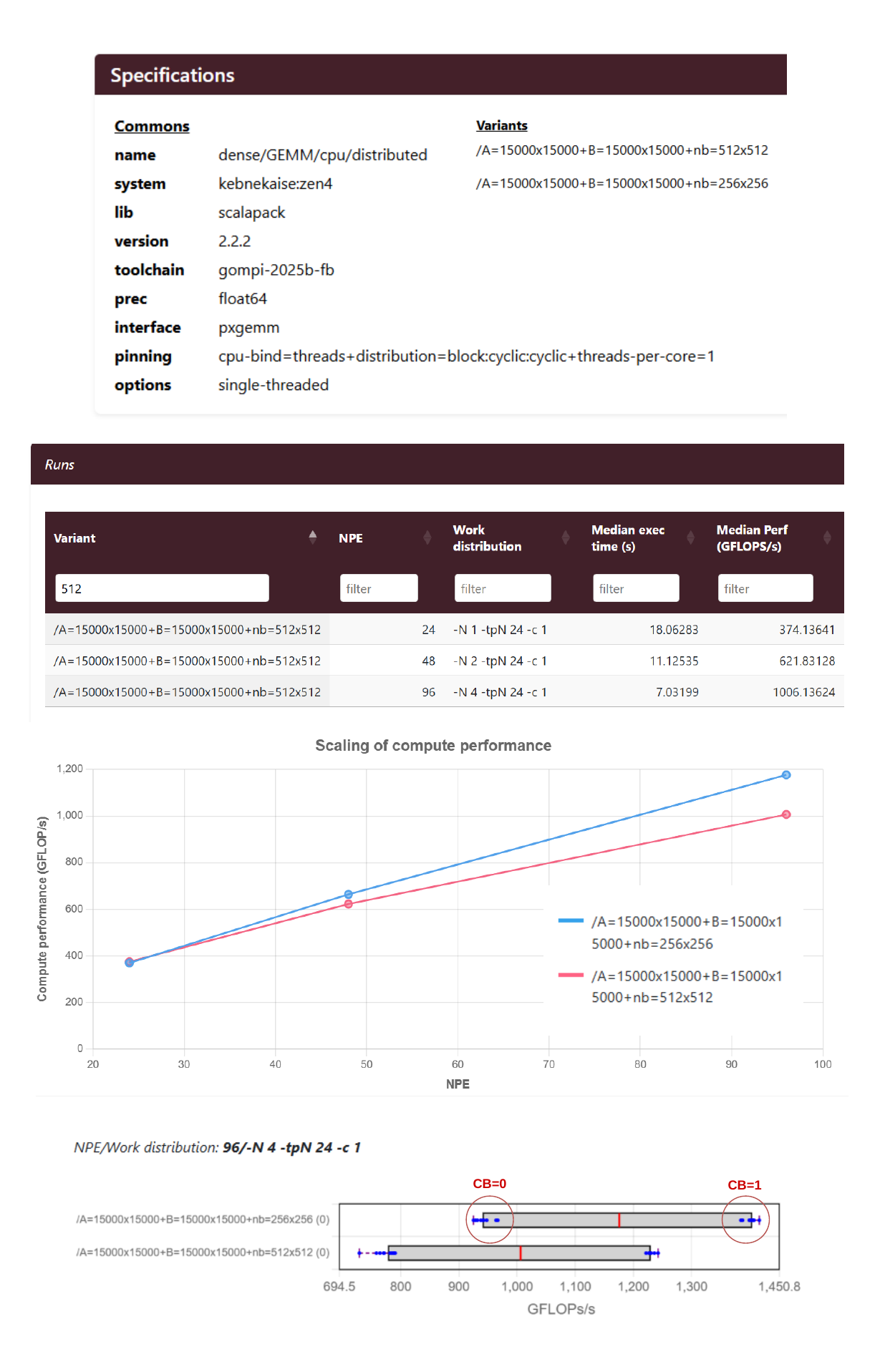}
    \caption{A snapshot of LAAB performance report for distributed DGEMM with square matrices of size 15000×15000 using ScaLAPACK 2.2.2 and the \texttt{gompi-2025b} toolchain on Kebnekaise
    system. The ``specifications" panel records the library installation version and toolchain, matrix size, precision, interface, and Slurm process pinning settings used. The report compares the effect of two block sizes---256 and 512---in the 2D block-cyclic distribution for distributed matrix multiplication. The ``runs" panel summarises execution time and performance at 24, 48, and 96 processing elements. The plot below the runs panel shows that the experiment with 512 block size scales less effectively than that with 256 block size. However, the distribution box plot at the bottom indicates that this difference in performance is not consistent; that is, it is possible that sometimes, the performance pertaining to 512 can be higher than that of 256.
    }
    \label{fig:laab-snapshot}
\end{figure*}

3) \textit{\textbf{Benchmark execution (JUBE).}} 
Each step is benchmarked for different parameter settings, such as library versions, NPEs and process-pinning configuration.
The benchmark runs are submitted to the HPC system through workload managers such as SLURM~\cite{yoo_slurm_2003} or Flux~\cite{ahn_flux_2018}, using batch scripts. Each submission is referred to as a job. We use JUBE to abstract the underlying workload manager and manage job execution. JUBE allows batch scripts to be parameterised, enabling the same scripts to be reused across different configurations. For every job, JUBE creates a self-contained run directory containing the generated batch scripts, parameter values, and the output data resulting from the execution. The source code is also copied into a sandbox directory before the job is run, ensuring that the exact code and configuration remain retrievable. 

To account for run-to-run variability, each job execution includes multiple benchmark measurements. However, when individual job executions are short, the influence of transient system effects, such as contention for shared resources, cannot be ruled out~\cite{hunold_reproducible_2016}. It is therefore important to repeat the job itself at different points in time and evaluate the measurements collectively. We refer to this repeated job execution over time as \textit{continuous benchmarking}. For example, consider the two box plots for the experiments shown in Fig.~\ref{fig:laab-snapshot}. Each box plot summarises measurements from two repeated job executions, corresponding to two continuous benchmark runs (CB=0,1), with each run itself comprising several repetitions. Owing to transient effects, the measurements within each experiment form two distinct groups, observed at two ends of the boxplots. 
% Therefore, to improve the reliability of the reports by providing transparency into transient effects, measurements are collected across multiple benchmark runs.

4) \textit{\textbf{Processing of benchmark measurements (laab-inspector)}.}
After each job is executed, its output log files are parsed to construct a performance profile for the step under evaluation. 
Constructing a performance profile may require output logs from multiple jobs---for example, computing speedup requires retrieving the logs of the job run with one processing element, while recomputing a ranking of library versions requires benchmark results from the other library versions.
For example, in Fig.~\ref{fig:gemm_bp}, each of the 10 library  versions is evaluated using four continuous benchmark runs. Generating the figure therefore requires logs from 10×4=40 job runs.
% For the experiment in Figure~\ref{fig:gemm_bp}, for instance, each library version is evaluated using four continuous benchmark runs, whose measurements are combined to construct the corresponding box plot. Therefore, generating Figure~\ref{fig:gemm_bp} requires gathering the output logs from 10×4=40 job runs across the ten library versions.
Then, our utility tool, \texttt{laab-inspector}, accepts a set of tags describing the run, retrieves performance profiles from related jobs, combines them with the newly generated profile, and recomputes the performance metrics. To identify performance regressions, the tool applies partial-ranking methods~\cite{sankaran_ranking_2025} to produce rankings that allow for ties in datasets. In Fig.~\ref{fig:gemm_bp}, the numbers in parentheses alongside the y-axis labels indicate the partial ranks assigned to each version using the method described in~\cite{sankaran_ranking_2025}.

The resulting performance profile, together with the self-contained run directories of the related jobs from JUBE are archived and stored in a repository. 

5) \textit{\textbf{Visualisation of performance reports (dashboards)}}.
The performance profiles are used to generate performance reports, which are visualised through a Flask-based web application that we refer to as the \textit{dashboard}.  If the archive containing the generated profiles is available, the dashboard can be invoked locally using the command \texttt{laab-dashboard <PROFILE\_DIR>}, with the path to the profile directory provided as input.  The index page of the dashboard lists the steps for which performance profiles are available.  Selecting one profile opens the performance report for that profile. Two or more profiles can be selected to generate a comparison report. 

 In general, the metrics included in a performance profile and the way in which the corresponding performance report is rendered may vary depending on the library type. To facilitate the development of customised dashboards, we developed \textit{Tvastar}~\cite{sankaran_laab-tvastar_2026}, which encapsulates reusable user-interface components for performance reporting, including filterable tables, box plots, and scaling plots. These components can be combined to construct dashboard layouts tailored to different reporting needs.

% and uses the \texttt{laab-inspector} component to enable interactive comparisons of user-selected profiles.

% The performance reports are presented through a Flask-based web application, which we refer to as \textit{dashboard}. The index page of the dashboard lists the steps for which performance profiles are available, and uses the \texttt{laab-inspector} component to enable interactive comparisons of user-selected profiles.
% The dashboard is deployed on a web server to provide shared access to centrally generated reports. Users can also run it locally to visualise reports produced from modified benchmarks.

6) \textit{\textbf{The transfer and retrieval of archives (laab-pigeon)}}. 
The dashboard can also be deployed on a web server to provide shared access to centrally generated reports. In this case, the archive containing the benchmark source code and performance profiles is transferred to a repository accessible to a web application that renders the performance report.
To this end, we developed \textit{Pigeon}~\cite{sankaran_laab-pigeon_2026}, a lightweight component that enables data transfer from a HPC system to a serving repository. It supports the implementation of sanity checks to prevent unintended or corrupt data from being included. It also 
%uses time-limited token-based access to prevent unauthorised data transfer, and 
allows users on an HPC system to retrieve the benchmark code required to reproduce a given step.

%The dashboard can be run either on a web server or locally in a user environment to visualise the reports for user modified benchmark. In both cases, 

\textbf{\textit{Workflow for HPC users and reviewers}}. 
Users and reviewers need not concern themselves with the internal components of the framework, such as setting up of JUBE, the inspectors, or Pigeon. Their starting point is the hosted dashboard, which presents performance reports for the available mathematical libraries. These reports are typically generated for predefined problem sizes. 
If the problem sizes or run-time options do not match the intended production workload, users can retrieve the minimal benchmark package—comprising the source code, Makefile, and job-submission script—from an HPC system to which they have access, modify the relevant settings, and re-run the benchmark. Performance profiles compatible with LAAB are then generated locally in the user space for the modified runs and can be viewed from a login node or laptop as described in point (5).

% running \texttt{laab-dashboard <path-to-profile-directory>}.

% The index page of the dashboard lists the available performance profiles. Selecting one profile opens the performance report for that profile. Two or more profiles can be selected to generate a comparison report. 

\section{Experiment}
% \textbf{\textit{Experiment}}. 
For demo, let us consider the multiplication of double-precision square matrices of size 15000×15000 with computations distributed across multiple processes using the ScaLAPACK library, which employs a two-dimensional block-cyclic distribution~\cite{choi_pumma_1994}.
The matrices are partitioned into small blocks that are cyclically assigned to the participating processes.
% \p{which processes? you only mentioned (a grid of) nodes}
The size of each block is a parameter of the computation and we want to compare the scaling performance obtained with block sizes of 256×256 and 512×512. 
% \p{for different number of nodes, right?} \as{I avoided the use of "Nodes" until here. I changed performance to scaling performance} 
The snapshot in Fig.~\ref{fig:laab-snapshot} shows the comparison report for this experiment. The benchmark was executed using ScaLAPACK 2.2.2 built with the \texttt{gompi-2025b-fb} toolchain on Kebnekaise. The suffix \texttt{fb} to the toolchain in Fig.~\ref{fig:laab-snapshot} indicates that ScaLAPACK is linked through the FlexiBLAS interface, with OpenBLAS selected as the backend. 
The experiment was run with NPE sizes of 24, 48, and 96, distributed across one, two, and four nodes, respectively. 
% \p{you talk about processes, nodes, and NPE -- what are the differences? Then, in the next sentence you also bring up tasks; What are those? We only talked about DGEMM.} \as{Until here, I only use the word "process". NPE is the number of processing elements. For this experiment, number of processes and NPE are equivalent, but it need not be so, e.g., if one process use more than one CPU cores.}
In the report, for each NPE, the work distribution is specified by the number of nodes (\texttt{-N}), number of processes per node (\texttt{-tpN}), and CPUs per processes (\texttt{-c}). The Slurm pinning setting indicates that the processes are filled for each node before moving to the next and are distributed cyclically across sockets, with one hardware thread per core. For this experiment, the scaling plot in the report shows that the use of 512×512 block size results in less effective scaling than the use of 256×256 block size. 

Recall that the performance metrics were obtained by summarising the results of two continuous benchmarks. The distribution plot indicates substantial transient variation between the two runs, possibly caused by shared resource contention, leading to measurements observed in two groups for each experiment. This variation is large enough that neither block size exhibits a consistent performance advantage, and the 512×512 block size occasionally outperforms the 256×256 variant. The report captures this behaviour, which can then be used for further investigation.
% rephrased.
% which can trigger a workflow for further investigation.

If this matrix-multiplication operation appears as a step in a production workload, its contribution to the total compute time can be estimated by multiplying the median execution time indicated in the report and the number of times that step is expected to be performed.
%LAAB allows the corresponding benchmark to be retrieved and rerun with a different problem size, if required.

\section{Conclusion}
This paper introduces LAAB, a framework for benchmarking and reporting the performance of computational steps provided by mathematical-library installations on HPC systems. 
% \p{on which system did you use LAAB?} \as{Kebnekaise and JSC systems, but I would call them tests and not deployments.} \p{ It's quite strong to be able to say "presently deployed at JSC...."} \as{I don't say that because we are not that strong yet, but to me, that is the goal.}
LAAB provides a space to address four objectives: (i) traceability of performance reports to the exact library installation and runtime configuration; (ii) compatibility between core-library and high-level software evaluations; (iii) reliable treatment of measurement variability for effective performance regression identification; and (iv) accessible reports that users can inspect, compare, and reproduce benchmarks. We described the LAAB components required to develop and operate the framework, presented a preview of the resulting performance reports and examples of insights they could provide.

The framework brings together EasyBuild and JUBE along with additional software components to make performance reports traceable and accessible. With LAAB, we aim to create an ecosystem 
that gathers contributions towards reliable performance reporting of mathematical operations across different application domains.
The performance reports are intended to provide HPC personnel, users, and reviewers with better insight into the performance of mathematical operations that constitute the building blocks of HPC applications. 
This work has presented an overview of LAAB and constitutes a part of a series of works; subsequent contributions will provide a detailed development guide, usage guidelines, and demonstrations of specific use cases.

% This work presents only an overview of LAAB. In subsequent work, we will elaborate on further benchmarks and use cases.

\textbf{\textit{Availability of code}}. The components of the LAAB framework are implemented as open-source projects and made available collectively through Zenodo~\cite{sankaran_laab-hpc_2026}. 

% \p{acknowledgments: HPC2N + eSSENCE funding. E.g.: "This research was conducted using the resources of the High-Performance Computing Center North (HPC2N), Sweden. This work was partially supported by the eSSENCE Programme under the Swedish Government's Strategic Research Initiative."}
% \as{added this in the first page footnote. Without the comments, only the "availability of code" fall outside the 8 page limit.}

\bibliography{reference-laab}
\bibliographystyle{IEEEtran}
\end{document}